\documentclass[epj,twocolumn]{webofc}
\usepackage[varg]{txfonts}
\usepackage{wasysym}
\usepackage{hyperref}
\woctitle{ISVHECRI 2016}

\begin{document}

\title{The STM32 microcontroller based pulse intensity registration system for the neutron monitor}

\author
{
\firstname{Alexander}
\lastname{Shepetov}
\inst{1,2}
\fnsep
\thanks{\email{ashep@tien-shan.org}}
\and
\firstname{Alexander}
\lastname{Chubenko}
\inst{1}
\and
\firstname{Olga}
\lastname{Kryakunova}
\inst{3}
\and
\firstname{Nikolay}
\lastname{Nikolayevsky}
\inst{3}
\and
\firstname{Nazyf}
\lastname{Salikhov}
\inst{3}
\and
\firstname{Victor}
\lastname{Yanke}
\inst{4}
}

\institute
{
{P.~N.~Lebedev Physical Institute of Russian Academy of Sciences (FIAN), 119234, Leninsky pr., 53, Moscow, Russia}
\and
{Tien~Shan Mountain Cosmic Ray Station, 480020, Mitina str 3b, Almaty, Kazakhstan}
\and
{Institute of Ionosphere, 050020, Kamenskoye plato, Almaty, Kazakhstan}
\and
{P.~N.~Pushkov Institute of Terrestrial Magnetism, Ionosphere and Radio Wave Propagation of Russian Academy of Sciences (IZMIRAN), 142190 Troitsk, Russia}
}

\abstract
{
We present the outlines of a new microcontroller based data acquisition system which is aimed for reliable operation in a typical cosmic ray particle registration experiment. The system supports connection of up to 16~input signals and ensures the following operation functionality: (1)~stable monitoring of the intensity of a digital pulse signal, or digitization of a continuous potential level with a low time resolution (typically, in the limits of 1--100~s); (2)~registration of a continuous high-resolution (up to 5--10~$\mu$s) time series of the intensity of input signal; (3)~synchronization of registered time series  with both external (physical) or local (program-based) trigger signal; (4)~possibility of an on-the-fly change of the whole configuration of informational system (both the combination and type of input signals, time resolution and sum duration of the time series measurements, trigger logic, etc) immediately in operation time through convenient communication with a plain text message in dialog mode. In particular, the considered system is applied now for a long-term, high precision measurement of the counting rate of neutron signals at the NM64 type neutron supermonitor of the Tien~Shan mountain cosmic ray station, with a real-time representation of the whole collected dataset in a WWW database.
}

\maketitle

\section{Introduction}
\label{intro}
The data acquisition system newly created at the Tien~Shan mountain cosmic ray station is made on the basis of a \mbox{STM32F407} type microcontroller unit (MCU) \cite{STM32F405} with the main purpose to ensure the stable registration of particle detector signals during a prolonged time period, in combination with flexible and simple control over the measurement process immediately in the time of its operation. The system is aimed to register the intensity of electric pulse signals which can arrive from various kinds of cosmic ray detectors; to make a real time analysis of measurement results in an automatic regime; and to store the resulting data for permanent keeping in a globally accessible database with a web oriented interface. Originally, the system was designed for operation with output pulses of neutron-sensitive proportional counters used in informational channels of the Tien~Shan 18NM64 neutron supermonitor \cite{ontienmonitor}, but it can be applied in other detector set-ups of similar kind. The system supports long-term, high-precision registration of the intensity of signal pulses on its inputs: it can operate continuously with an uninterruptable duty cycle of up to some years, and permits to measure the intensity variations of the input signal with relative accuracy of up to 0.05--0.1\%. After being switched on, the system operates fully autonomously, and does not require any special service from the side of qualified personnel.

The basic functionality of the considered system consists of continuous measurement of the counting rate separately for each neutron detector of the monitor, of automatic analysis of the monitor operation stability, and of regular sending of accepted results into some external database. The typical duration of the pulse number counting time in these {\it monitoring} type measurements is of the order of 1--100~s. (In the world wide net of the cosmic ray variations registration that the Tien~Shan neutron supermonitor belongs to,  a 1~minute periodicity of intensity measurements  is historically accepted).

Besides the registration of input pulses which come immediately from the neutron monitor, the data acquisition system ensures the detection of the so called neutron multiplicity events, i.e. arrival cases of  a number of neutron signals within a short time (typically, of millisecond order). These events are caused by the momentary generation of a large number of evaporation neutrons inside the monitor as the  interaction result of an energetic cosmic ray hadron. Produced neutrons are living and diffusing within the monitor during a rather considerable time, being partially registered in this period by the monitor counters. The number of detected neutron pulses (called {\it multiplicity}), is proportional to the amount of generated neutrons, and the latter, in turn, depends on the energy of the primary cosmic ray interaction, since the average number of evaporation neutrons is a power function of this energy. Hence, registration of multiplicity events gives an opportunity to study the variations of the energy spectrum of cosmic radiation \cite{onmulti}.

The control program of the microcontroller unit forms the multiplicity events internally by analyzing the current flow of input pulses in real time in accordance with (arbitrary) rules which can be built in into its embedded program code. For compatibility with previous data series, 6~types of multiplicity signal are currently generated at the Tien~Shan monitor for the cases where 1, 2, 3, 4--5, 6--7, or 8--10 neutron pulses have been accepted from six counters of a standard NM64 type supermonitor unit within a 800~$\mu$s long gate time. If needed, the formation algorithm of the neutron multiplicity signal can be easily changed just in measurement time through re-programming of the MCU.

When using the considered system for   the data acquisition from some specialized cosmic ray particle detectors it can be necessary to register coincidences between the signals which come to inputs of various informational channels. (The prototype installation of this kind is a particle telescope like set-up consisting of a number of spaced scintillators separated by an absorber which permits to estimate the energy and direction of penetrative cosmic ray particles---muons \cite{dorman_tele}). In the same manner as for multiplicity events, the microcontroller program permits to solve this task in parallel to basic intensity measurements by forming the coincidence signal internally from the current flow of input pulses. At that, it is possible to change promptly any parameters of the operation algorithm (combination of input signals to form coincidence from, the resolution time, etc) immediately at the time of measurements through easy MCU re-programming. The existing program version ensures the registration of pairwise coincidences between two opposite scintillation counter layers, with 8 detectors in each layer, and with arbitrary combination of detector pairs for the muon telescope of IZMIRAN.

Both the formation of multiplicity events and tracing of coincidence signals are realized within the MCU in parallel and simultaneously with the basic pulse counting process.

One of the directions of the high energy cosmic ray investigation held at the Tien~Shan mountain station is connected with rather rare events when the core of a $10^{15}-10^{17}$~eV extensive air shower (EAS) passes in the  immediate vicinity ($0-5$~m) to the  neutron monitor. Sometimes, the events of this type are accompanied with instantaneous generation of an extremely large number of evaporation neutrons (of up to some thousands) as an interaction result of energetic EAS core hadrons with monitor substances; later on, these neutrons are quickly thermalized and diffuse in the monitor during some milliseconds long life time \cite{jopg2001}. The MCU acquisition program permits to trace the events of such a type in the current signal flow, and to register  precisely the time behavior of neutron intensity with a 10--30~$\mu$s time resolution uninterruptedly over the period of some milliseconds. The latter functionality, again, goes on in parallel and independently on the basic counting rate measurements. Switching on the tracing of EAS events,  the  configuration of registered temporal distributions (the list of informational channels which participate in EAS events formation, the resolution time, sum duration of distribution, etc) can be set immediately during the measurements by corresponding commands sent to the MCU.

Besides EAS events formed internally by the microcontroller itself, the high resolution time registration can be forced at any arbitrary moment by the arrival of a special control signal---the trigger, which can be elaborated outside the system, and comes as an electric pulse to a special input pin of the microcontroller. Such a synchronization type can be useful, for example, for the  registration of the signals from some outside neutron and gamma-ray detectors which should operate simultaneously and in parallel with the neutron monitor. The outer trigger synchronization mode can be switched on or off, again, by demand at any arbitrary time moment.

\section{Hardware design}
\label{sec-1}

\begin{figure}
\begin{center}
\includegraphics[width=0.5\textwidth, clip, trim=10mm 70mm 10mm 60mm]{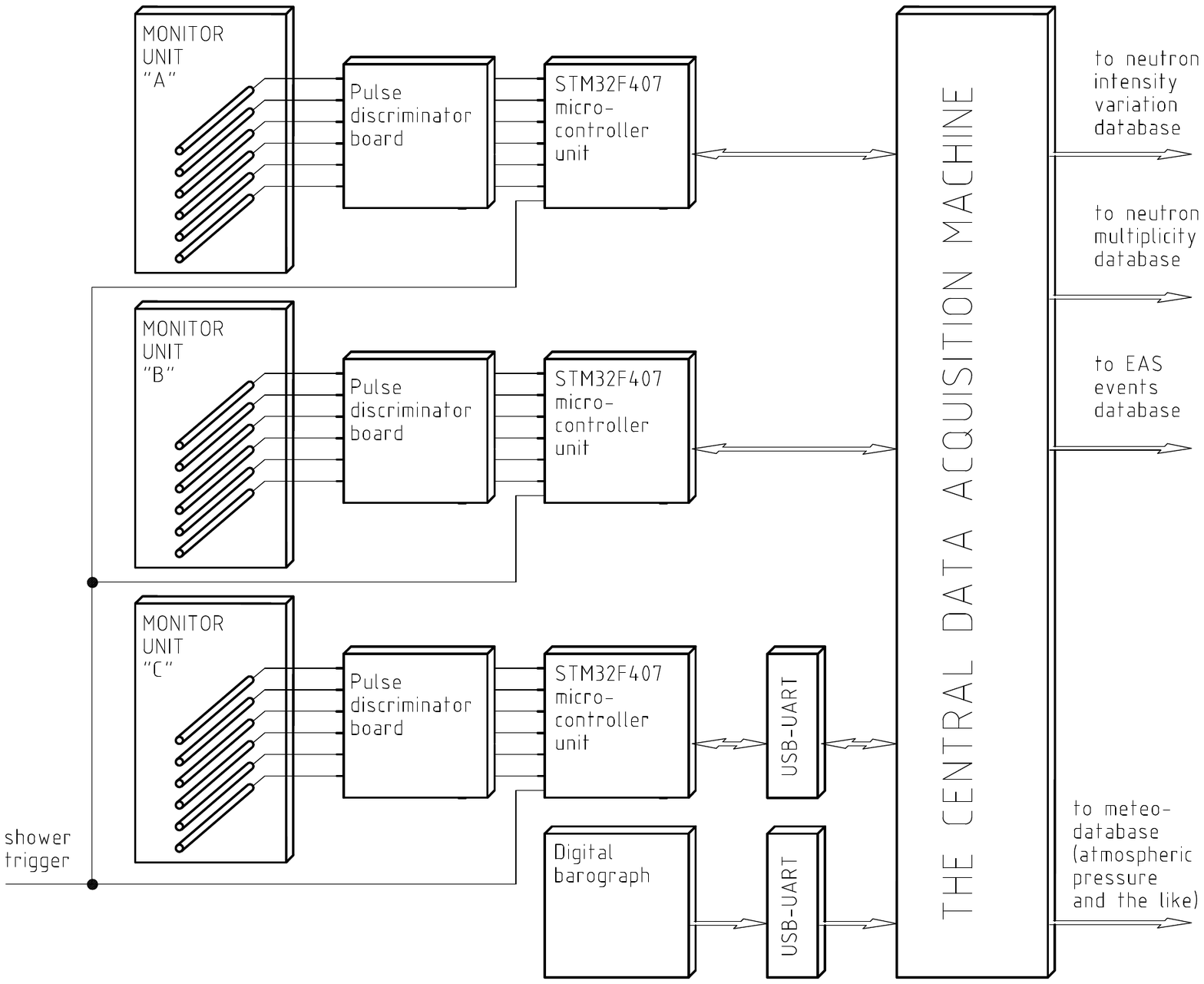}
\end{center}
\caption {Block scheme of the cosmic ray intensity data acquisition system used at the Tien~Shan 18NM64 neutron supermonitor.}
\label{neutron-functi}
\end{figure}

The further description of the MCU based data acquisition system is made on the example of its variant which has remained in continuous operation at the 18NM64 neutron supermonitor of the Tien~Shan mountain cosmic ray station since the beginning of 2015. The general block scheme of this system is shown in Figure~\ref{neutron-functi}.

\begin{figure*}[ht]
\begin{center}
\includegraphics[width=1\textwidth, clip, trim=10mm 30mm 10mm 23mm]{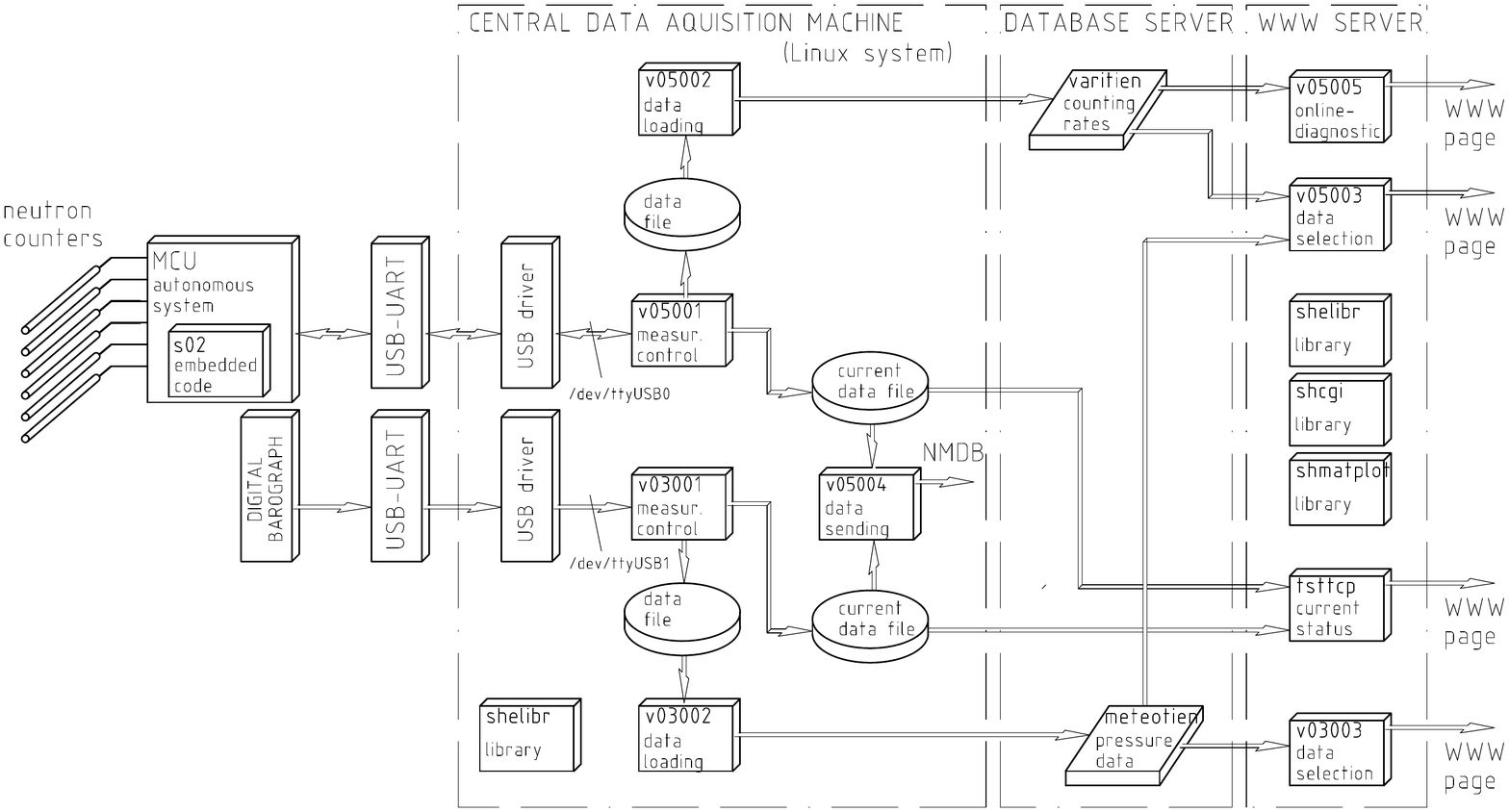}
\end{center}
\caption{General schematic of the Tien~Shan neutron supermonitor control program complex.}
\label{functiprg}
\end{figure*}

The Tien~Shan neutron supermonitor consists of three separate units, each of which has the dimensions of 2~$\times$~3~m$^2$ and includes six big, \diameter0.15~m~$\times$~2~m proportional neutron counters \cite{carmichel_supermonitor, hatton_supermonitor_inbook}. Output pulses with typical amplitude of $5-20$~mV and duration varying in the limits of $0.5-2$~$\mu$s are taken from the anode wire of each counter and transmitted through a transistor amplifier into a short shielded coaxial cable which connects this counter with corresponding informational channel on the data acquisition board. The latter board is located in immediate vicinity to monitor unit, so that the length of connection cables does not exceed $0.5-2$~m, and includes a set of six pulse discriminators necessary to properly form the neutron signals, as well as a processor module with a \mbox{STM32F407} type MCU installed. Pulse signals from discriminator outputs, which have a constant $\sim$3~V amplitude and a fixed 1~$\mu$s time duration, come to the MCU input pins for further operation. The high voltage of the neutron counters feeding, transmission coefficients of transistor amplifiers, and discriminator thresholds are individually tuned in such a way to ensure  for every counter its operation in beginning part of the plateau of its counting characteristic.

Physically, each discriminator and MCU carrier board comprises a separate device---an autonomous data registration system. Presently, there are three such systems installed at the Tien~Shan neutron supermonitor, each one  serving its own monitor unit.

All three microprocessor systems operate under the control of the one and the same embedded program code which provides complete internal functionality of the pulse signal operation described in previous section. Besides this, the embedded MCU code ensures the full duplex dialog-like interaction between autonomous measurement control system and the central data acquisition machine. The latter is generally responsible for sampling the incoming neutron intensity data, for their primary operation, and for loading of all measurement results into a permanent database, as well as for synchronous launching of the measurement process on all autonomous systems and continuous steady control over their current status during the whole operation time.

By the means of either an universal asynchronous receiver-transmitter (UART) module, or an USB node both of which are in-built into the \mbox{STM32F407} microprocessor, the autonomous registration system can interact with central control computer via a serial interface line. Using an UART based variant is preferential in the case when the length of the communication cable to autonomous system must be above 1.5--2~m (which is the case of monitor unit {\it C} in Figure~\ref{neutron-functi}), so the use of additional transistor amplifiers of the interface signal with a further UART$\leftrightarrow$USB converter at the side of central computer is desirable; practically, the distances of up to 30--50~m between the disposition of a detector system and control machine center can be achieved in this way without any significant difficulty. The whole information exchange goes in the form of a plain text messages, so there is no need for any specialized communication software, and every commonly used serial interface program is sufficient enough in the simplest testing case to send the commands to autonomous system, and to check back its reaction.

\section{Software complex}
\label{sec-2}

The block scheme of the complete program complex  presently used at the data acquisition system of  the Tien~Shan neutron supermonitor is shown in Figure~\ref{functiprg}. As it is seen in this figure, the program set consists of the following main parts.

Each autonomous data registration board which the outputs of the neutron counter channels are immediately connected to operates under the control of a specialized embedded program {\it s02} which plays the role of the MCU operation system, and ensures its proper reaction on asynchronous program interrupts caused by the arrival of input pulse signals. For compatibility with other use, the version of the embedded program  presently applied in the microprocessor registration systems of the Tien~Shan supermonitor supports the connection of up to 16 separate signals to every autonomous board, though only 6 of them are now in use by each MCU. The embedded program {\it s02} is created on the basis of the free ARM architecture microcontroller programming library {\it libopencm3} \cite{libopencm3} which provides the proper low-level access to the whole built-in periphery of the  \mbox{STM32F407} microprocessor box. The program is written in C programming language and converted into internal MCU machine code with the aid of a free open source toolchain from the {\it gcc-eabi} project \cite{gcc-arm-embedded}.

The central computer that all autonomous measurement
boards interact with works under the control of the free operation system \mbox{Linux}, which gives the possibility of both initiation of specialized programs to accept the data from autonomous registration systems, and of necessary network interaction with outer host machines: some database server to store the measurement results, the network time protocol ({\it nntp}) server to keep the constant synchronization of internal computer time, and an {\it http} server to present the current status of running measurements for remote observation through the web.

Normally, there is a specialized supervising program, {\it v05}, which supports operative control over the measurement process from the side of the central machine. It is this program which sends the commands for configuration of the autonomous system at  the beginning of the measurements, keeps a steady check over its operation afterwards, accepts the results from the  autonomous system, and saves these data temporarily in a local disk file. Simultaneously, there are just three separate {\it v05001} program instances  constantly operating on the central machine, each of which controls the measurements of its own autonomous system. Periodically, according to timer signals another {\it v05002} type process is initialized which sends the whole information stored locally to the outer database server.

Another two programs, {\it v03001/v03002}, operate in an analogous fashion, and are aimed to track the atmospheric pressure which is necessary for proper correction of the raw neutron counting rate data, and is continuously provided by a digital barograph.

The general database of  the Tien~Shan mountain cosmic ray station is used as a permanent storage for  information which comes from all experiments which are currently active at the station. At the present time the database functionality is supported by the {\it PostgreSQL} server program \cite{postgres}, and an access to different types of data kept there is provided through specialized script programs launched from the common WWW~page of the Tien~Shan station \cite{tieneng}. Clicking on the  corresponding reference in this page the WWW~interface programs {\it v05003} and {\it v03003} can be initiated which select the necessary information on the cosmic ray intensity measured by neutron counters of the Tien~Shan supermonitor, and present it on the page in graphical or textual forms. A sample of such an output is shown in  Figure~\ref{intensfig}.

Besides the storage of collected neutron monitor information in the local database of the Tien~Shan station, the {\it v05/v03} programs can re-send these data immediately during the measurements to the outer European Real-time Neutron Monitor Database (\mbox{NMDB}) \cite{nmdb}. 
The special program script {\it v05004} which is initiated automatically after completion of every minutely measurement is responsible for proper communication with this remote server, and for timely data transmission.

High level program components of the {\it v05/v03} data manipulation complex are written in the interpreted programming language Python to ensure good portability between various operation environments which could be met at different server machines, as well as simplicity and transparency of program code from the programmer's point of view. For their operation all these programs use a number of auxiliary procedures defined in a separate library module {\it shelibr}; besides, the programs responsible for data representation on a WWW page utilize specific means of web server communication from the  {\it shecgi} module, and {\it shematplot} convenience envelope over the free Python language graphic library {\it matplotlib} \cite{matplotlib}.

\section{Conclusion}
\label{sec-3}

\begin{figure}
\begin{center}
\includegraphics[width=0.45\textwidth, trim=0mm 0mm 0mm 0mm]{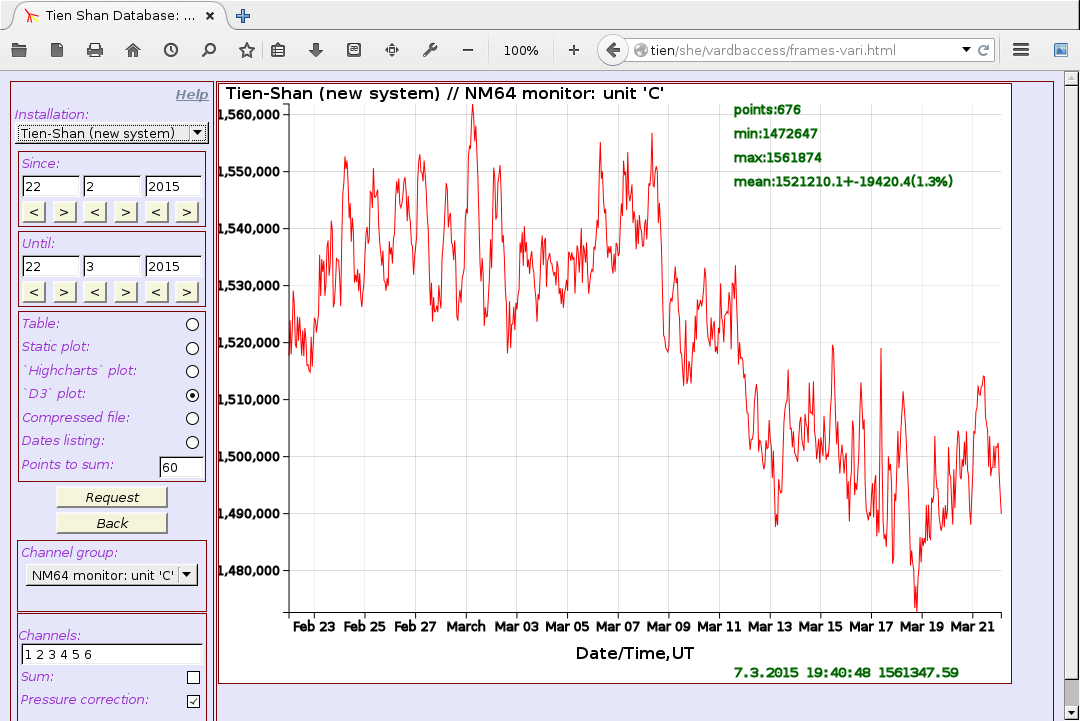}\\
\includegraphics[width=0.45\textwidth, trim=0mm 0mm 0mm 0mm]{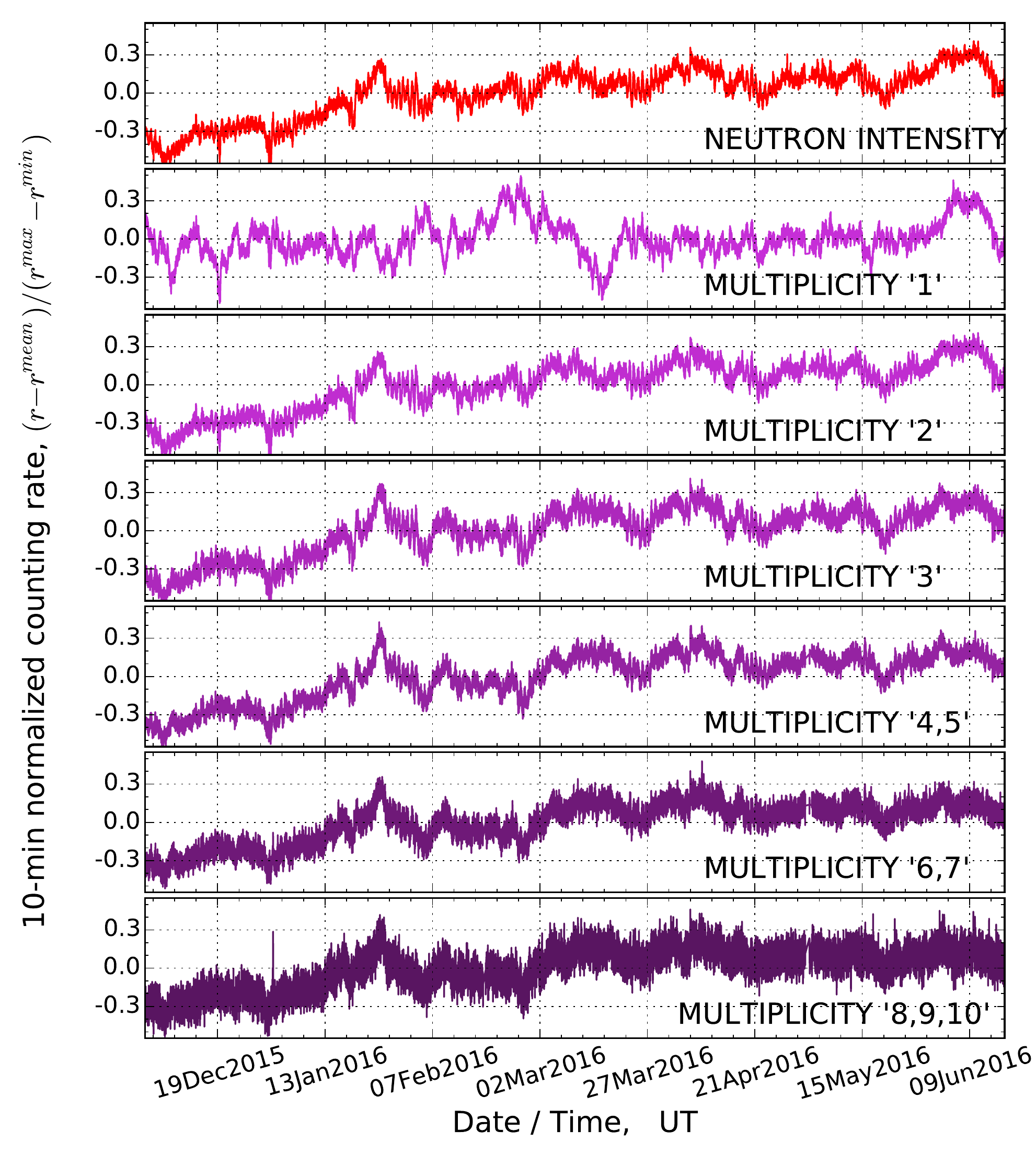}\\
\includegraphics[width=0.45\textwidth, trim=0mm 0mm 0mm 0mm]{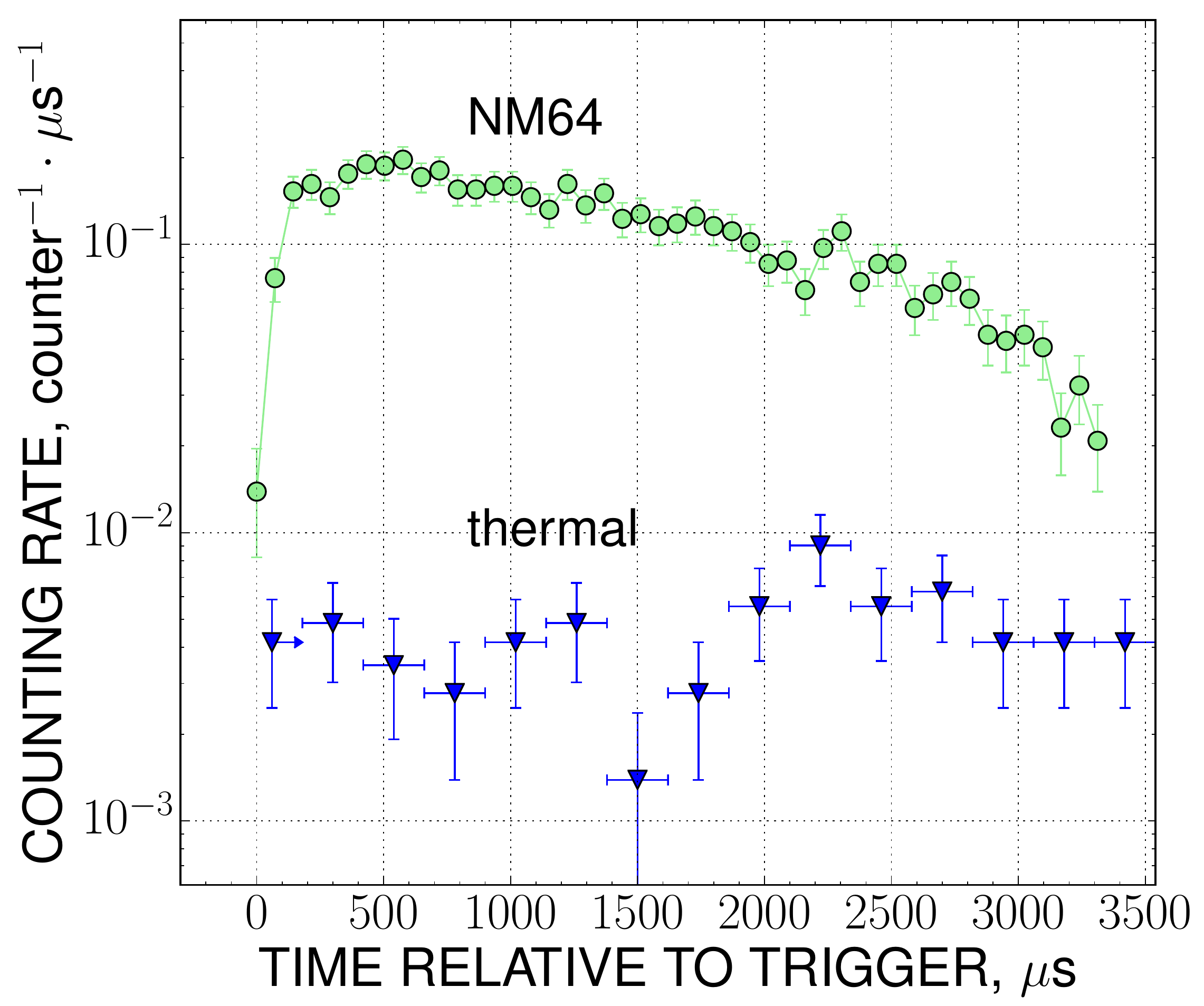}
\end{center}
\caption {A sample of cosmic ray data from the Tien~Shan neutron supermonitor. Top panel: representation of the current neutron counting rate measurements on the WWW site page in graphic form. Middle panel: the half-year long uninterruptable record of the neutron counting rate, and the intensity of neutron multiplicity events taken from archive database and presented with 10~min time resolution; all data are corrected for variations of the atmospheric pressure. Bottom panel: high-resolution (70~$\mu$s) time distributions of pulse intensity in the neutron counters of a NM64 supermonitor unit, and in a nearby thermal neutrons detector by the close passage of an EAS core.} 
\label{intensfig}
\end{figure}

During a 3~year long period of continuous exploitation of the \mbox{STM32F407} MCU based data acquisition system at the Tien~Shan neutron supermonitor it has demonstrated good operation reliability, and high stability of precision measurements of the input pulse intensity, as well as sufficient convenience for practical use in cosmic ray experiments. Some typical results of this work are illustrated by the plots in Figure~\ref{intensfig}.

The use of a cheap but sufficiently powerful modern MCU in the development of  cosmic ray experimental set-ups opens the way for its further modernization and for a noticeable increase in the  comprehension of the resulting measurement data without the need of any considerable redesign of existing devices. Hence, an easy replacement of the input pulse counting module in the above described microcontroller driver program {\it s02} to an analogous one which can interact with the built-in analog-to-digital converter (ADC) node of MCU,  leaving alone all other functional modules of this program (those of the USB and UART interface support, data storage facility, outer command interpreter, etc) permits to apply the same autonomous registration board in the role of a fast  multichannel ADC system. This option is applied at the Tien~Shan cosmic ray station for developing various data sampling devices primarily intended for the detection of continuous electric signals (the radio- and optic emission sensors, electric field measurements, micropulsation of atmospheric pressure, and the like). By this, the feature of a dialog-based textual control over configuration parameters set of the {\it s02} driver program is very convenient for a fast and flexible tuning of the whole data registration process to specific needs of every particular detector.

It should be noted that the use of compact microprocessor technique with serial communication through a USB interface line and low power consumption allows for the creation of portable data registration set-ups for various mobile autonomous detector systems (such as the distant EAS particle detectors \cite{ontien-nim2016}, remote sensors of the optic and gamma-ray radiation from thunderclouds \cite{thunderour2016}, seismological stations \cite{undgmouacou}, and the like),  which are supposed to be used widely at the Tien~Shan detector complex. 

%
%

\end{document}